\documentclass[usenatbib,fleqn]{mnras}

\usepackage{aas_macros}
\usepackage{amsmath}
\usepackage{amssymb}
\usepackage{booktabs}
\usepackage{bm}
\usepackage{color}
\usepackage{textcomp}
\usepackage{gensymb}
\usepackage{graphicx}
\usepackage{hyperref}
\usepackage{lineno}
\usepackage{siunitx}
\usepackage{stackengine}
\usepackage{verbatim}
\usepackage{lineno}

\usepackage [english]{babel}
\usepackage [autostyle, english = british]{csquotes}
\MakeOuterQuote{"}

\newcommand{\xfigure}[1]{
\begin{center}
\includegraphics[width=0.5\textwidth]{figs/#1}
\end{center}
}

\newcommand{\yfigure}[1]{
\begin{center}
\includegraphics[width=0.9\textwidth]{figs/#1}
\end{center}
}

\begin{document}
\title{The PAU Survey: Photometric redshifts using transfer learning from
simulations}
\author[PAUS deep learning redshifts]{M. Eriksen$^{1}$\thanks{E-mail: eriksen@pic.es}\thanks{Also at Port d'Informaci\'{o} Cient\'{i}fica (PIC), Campus UAB, C. Albareda s/n, 08193 Bellaterra (Cerdanyola del Vall\`{e}s), Spain},
 A. Alarcon$^{2}$,
 L. Cabayol$^{1}$,
 J. Carretero$^{1}\footnotemark[2]$,
 R. Casas$^{3,4}$, 
 \newauthor
 F. J. Castander$^{3,4}$,
 J. De Vicente$^{5}$,
 E. Fernandez$^{1}$,
 J. Garcia-Bellido$^{6}$,
 \newauthor
 E. Gaztanaga$^{3,4}$,
 H. Hildebrandt$^{7}$,
 H. Hoekstra$^{8}$,
 B. Joachimi$^{9}$,
 R. Miquel$^{1,10}$,
 \newauthor
 C. Padilla$^{1}$,
 E. Sanchez$^{5}$,
 I. Sevilla-Noarbe$^{5}$,
 P. Tallada$^{5}$\footnotemark[2] \\
$^{1}$ Institut de F\'{\i}sica d'Altes Energies (IFAE), The Barcelona Institute of Science and Technology, 08193 Bellaterra (Barcelona), Spain \\
$^{2}$ HEP Division, Argonne National Laboratory, Lemont, IL 60439 \\
$^{3}$ Institute of Space Sciences (ICE, CSIC), Campus UAB, Carrer de Can Magrans, s/n, 08193 Barcelona, Spain \\
$^{4}$ Institut d'Estudis Espacials de Catalunya (IEEC), E-08034 Barcelona, Spain \\
$^{5}$ Centro de Investigaciones Energ\'eticas, Medioambientales y Tecnol\'ogicas (CIEMAT), Avenida Complutense 40, 28040 Madrid (Madrid), Spain \\
$^{6}$ Instituto de F\'{\i}sica Te\'orica (IFT-UAM/CSIC), Universidad Aut\'onoma de Madrid, 28049 Madrid, Spain \\ 
$^{7}$ Ruhr-University Bochum, Astronomical Institute, German Centre for Cosmological Lensing, Universitätsstr. 150, 44801 Bochum, Germany \\ 
$^{8}$ Leiden Observatory, Leiden University, Niels Bohrweg 2, 2333 CA, Leiden, The Netherlands \\
$^{9}$ Department of Physics and Astronomy, University College London, Gower Street, London WC1E 6BT, UK \\
$^{10}$ Instituci\'o Catalana de Recerca i Estudis Avan\c{c}ats (ICREA), 08010 Barcelona, Spain
}

\maketitle

\begin{abstract}
In this paper we introduce the \textsc{Deepz} deep learning photometric
redshift (photo-$z$) code. As a test case, we apply the code to
the PAU survey (PAUS) data in the COSMOS field. \textsc{Deepz} reduces the $\sigma_{68}$ 
scatter statistic by 50\% at $i_{\rm AB}=22.5$ compared to existing
algorithms. This improvement is achieved through various methods, including
transfer learning from simulations where the training set consists of
simulations as well as observations, which reduces the 
need for training data. The redshift probability
distribution is estimated with a mixture density network (MDN), which produces 
accurate redshift distributions. Our code includes an autoencoder to reduce noise and extract features
from the galaxy SEDs. 
It also benefits from combining multiple networks, which lowers the
photo-$z$ scatter by 10 percent.
Furthermore, training with randomly constructed coadded
fluxes adds information about individual exposures, 
reducing the impact of photometric outliers. In addition to opening
up the route for higher redshift precision with narrow bands, 
these machine learning techniques can also be valuable for broad-band surveys.
\end{abstract}


\section{Introduction}
Galaxy surveys provide invaluable information for a wide set of science applications.
They enable a census of the galaxy population and can constrain cosmological
models \citep{Gaztanaga2012, Weinberg2012, Eriksen2015},
where the galaxies act as tracers of the underlying dark matter 
field or are used to measure weak gravitational lensing
\citep{Bartelmann2001, Hoekstra2008}.
There are two main types of galaxy surveys: spectroscopic and photometric. 
Spectroscopic surveys have high redshift precision, but for limited galaxy
samples. Photometric broad band surveys cover larger volumes and
fainter galaxies, but their redshift precision is much lower \citep{Baum1962,Koo1985, Benitez2000, Hildebrandt2010, Salvato2019}.

The redshift precision of broad-band surveys is limited
by their filter width. An alternative approach is to use narrow-band imaging to obtain
high precision redshift estimates for a large sample of galaxies. 
The Physics of the Accelerating Universe Survey (PAUS)
 implements this idea using 40 narrow bands spaced uniformly in
the optical wavelength range from 4500\AA\ to 8500\AA\ \citep{Padilla2019}. This 
higher wavelength
resolution allows for detecting more features in the spectral energy
distribution (SED), leading to a better redshift determination
\citep{Marti2014, Eriksen2019}. For $i_{\rm AB} < 22.5$, \citet{Eriksen2019} 
demonstrated that PAUS attains its intended precision, reaching $\sigma_z = 0.0037 (1+z)$ for a selected 50\% of galaxies 
with secure spectra in zCOSMOS DR3 \citep{Lilly2007}. This precision is about a
magnitude better 
than with a typical broad-band survey.

The redshift estimates by \citet{Eriksen2019} were derived with \textsc{BCNz2}, a template based photometric
redshift code tailored to achieve high precision redshifts with PAUS. This
code used a linear interpolation between continuum spectral energy density (SED), 
added additional emission lines and also fitted for zero-points. A global 
zero-point was determined per band, while the code additionally allowed for a free scaling 
between the broad and narrow bands per galaxy.
The use of a template based code was chosen for two reasons. Initially we needed to derive the redshift
for samples of hundreds of galaxies, which are insufficient for
training. Furthermore, previous tests on machine
learning (ML) codes on simulations had not managed to achieve
the target PAUS photo-$z$ precision with a realistic training sample.

Despite theoretically being a versatile method, the \textsc{BCNz2}
template fitting code is hard to extend in different directions (appendix \ref{bcnz_description}). For example, 
the non-linear minimisation was difficult to combine with a model where the 
individual emission line strengths were varying with correlated priors between
the lines. Other difficulties included extending the statistical fitting to also 
account for photometric outliers (appendix \ref{effect_outliers}) or efficiently including priors on the different galaxy
types during the minimization. Also, formally one should estimate the redshift by integrating over the space of linear SED combinations and
not only consider the 
minimum (Alarcon in prep.). Together with other difficulties, technical issues
have made the template fitting approach hard to develop further. In this paper, 
we instead investigate applying machine learning techniques to determine PAUS redshifts. 

Machine learning redshift determination has a long history, with the \textsc{ANNz}
\citep{Collister2004} neural network code being one of the earliest examples. 
Furthermore, there are many codes, implementing common 
machine learning algorithms like
neural networks (\textsc{Skynet}) \citep{Bonnett2015}, support vector
machines (\textsc{SpiderZ}) \citep{Jones2017} and tree based
codes (\textsc{tpz}) \citep{Kind2013}. 
Machine learning codes offer certain advantages over template fitting methods. Since the machine learning
methods directly map magnitudes and/or colours to redshifts, one is not required to
model the SEDs, which can be challenging at high redshifts. For PAUS, the accurate SED modelling started to become a 
potential limitation for the high redshift precision target. Furthermore, the
direct colour-redshift mapping makes the model insensitive to global zero-points.

Constructing the training sample is a central problem to
estimate photometric redshifts with machine learning. This sample has been built from precise redshift information from spectroscopic surveys,
e.g. zCOSMOS \citep{Lilly2007} 
or VIMOS VLT Deep Survey (VVDS) \citep{Fevre2005}. These spectra are also required to cover the
colour space \citep{Masters2015}, sampling different types of galaxies. These limited training sets already pose serious problems for broad-band photo-$z$
and become a challenge for a magnitude better photo-$z$ precision that PAUS aims to achieve.

Transfer learning is an approach for reducing the requirement
on the training sample \citep{Pan2010}. Instead of training the network from scratch, one
can start training a network which has previously been trained on different data. The 
network can even benefit from using networks trained on quite different data.
In this paper, we focus on simulations, that resemble the observations. Combining the simulations and data has the
ability of reducing the need for training data. While attempted in
various forms (e.g. \citealt{Vanzella2004,Hoyle2015}), it is not commonly used. 

Machine learning techniques can be divided into different categories. The
most widely used  is supervised learning, which compares a prediction with a label 
(truth value). Even with dedicated surveys, redshift measurement of the 
faintest galaxies is considered  time consuming \citep{Masters2019}. 
These surveys usually include tens to hundreds of thousands of spectra for
specific targets. By contrast, e.g. the Dark Energy Survey (DES) and 
Kilo-Degree Survey (KiDS) offer
hundreds of millions of galaxies to $i_{\rm AB} < 24$, with photometric
information. In this paper we study the use of  autoencoders, which can be used without
knowing the redshift (unsupervised)
and has the potential advantage of potentially being able to 
train using a million of galaxies from PAUS.

This paper is built up in the following manner. First, \S \ref{data_architecture} describes the PAUS data, the network architecture
and the training procedure. In \S \ref{transfer_learning} we study the usage of transfer
learning from simulations. 
Then \S \ref{auto_encoders}
shows how autoencoders can be used to reduce the noise. 
Later in \S \ref{individual_exposures} we develop and
test a method for including individual exposures.
In \S \ref{validation} we 
validate the redshift probability distributions and introduce quality cuts, and we summarise and
conclude in \S \ref{conclusions}.

\section{Deep learning photometric redshifts}
\label{data_architecture}
This paper uses the same
input data as \citet{Eriksen2019} (\textsc{BCNz2}) and 
\citet{BKGnet}. For completeness, 
\S \ref{input_data} briefly describes the PAUS data, the external broad bands and the
spectroscopic catalogue. In \S \ref{network_architecture} we describe the network 
architecture, in \S \ref{pred_prob_func} the mixture density network to estimate the
redshift distributions and in  \S \ref{training_procedure} the training procedure.

\subsection{Input data}
\label{input_data}
This paper focuses on the data from the 
Cosmological Evolution Survey (COSMOS)
field\footnote{http://cosmos.astro.caltech.edu/} where we have PAUS observations and there are
abundant spectroscopic measurements. The COSMOS field also has a large set of photometric surveys, 
covering the wavelength range from ultra-violet to infrared. Our 
fiducial setup uses the 
Canada-France-Hawaii Telescope Lensing Survey
(CFHTLenS) $u$-band and the $B,V,r,i,z$ bands from the Subaru telescope as in
\citet{Eriksen2019}. As the spectroscopic catalogue, we use 8566 secure
($3 \leq \rm{CLASS} \leq 5$)
redshifts from the zCOSMOS DR3 survey \citep{Lilly2009} that are observed with all 40 narrow bands.

The PAUS data are acquired at the William Herschel Telescope (WHT) with the
PAUCam instrument and transferred to the
Port d'Informació Científica (PIC,  \citealt{Tonello2019}). First the images are detrended in the \textsc{nightly} 
pipeline (Serrano et al. in prep). Our astrometry is relative to {\it Gaia} DR2 \citep{Brown2018}, while the photometry is calibrated relative to the Sloan
Digital Sky Survey (SDSS) by 
fitting the Pickles
stellar templates \citep{Pickles1998} to the $u,g,r,i,z$ broad bands from 
SDSS
\citep{Smith2002} and then predicting the expected fluxes in the narrow bands. The
final zero-points are determined by using the median star zero-point for each
image.

PAUS observes weak lensing fields (CFHTLenS: W1, W3 and W4) with deeper broad-band data from external surveys. PAUS uses forced photometry, assuming known galaxy positions, morphologies and sizes from external catalogues.
The photometry code
determines for each galaxy the radius needed to capture a fixed fraction of
light, assuming the galaxy follows a Sérsic  profile convolved with a known Point 
Spread Function (PSF). The algorithm uses apertures that measure 62.5\% of 
the light, since this is considered statistically optimal. A given galaxy is observed several times (3-10) from different overlapping exposures. The coadded fluxes 
are produced using inverse variance weighting of the individual measurements. 
As described in \S \ref{individual_exposures}, we also train the network using 
individual fluxes.

\subsection{Network architecture}
\label{network_architecture}
For a reminder of the basics of neural networks, we refer the reader to \citet{LeCun2015}. Moreover, Appendix \ref{deep_basics} provides 
some basics on 
neural networks and introduces the terminology used in this paper.

\begin{figure}
\includegraphics[width=0.46\textwidth]{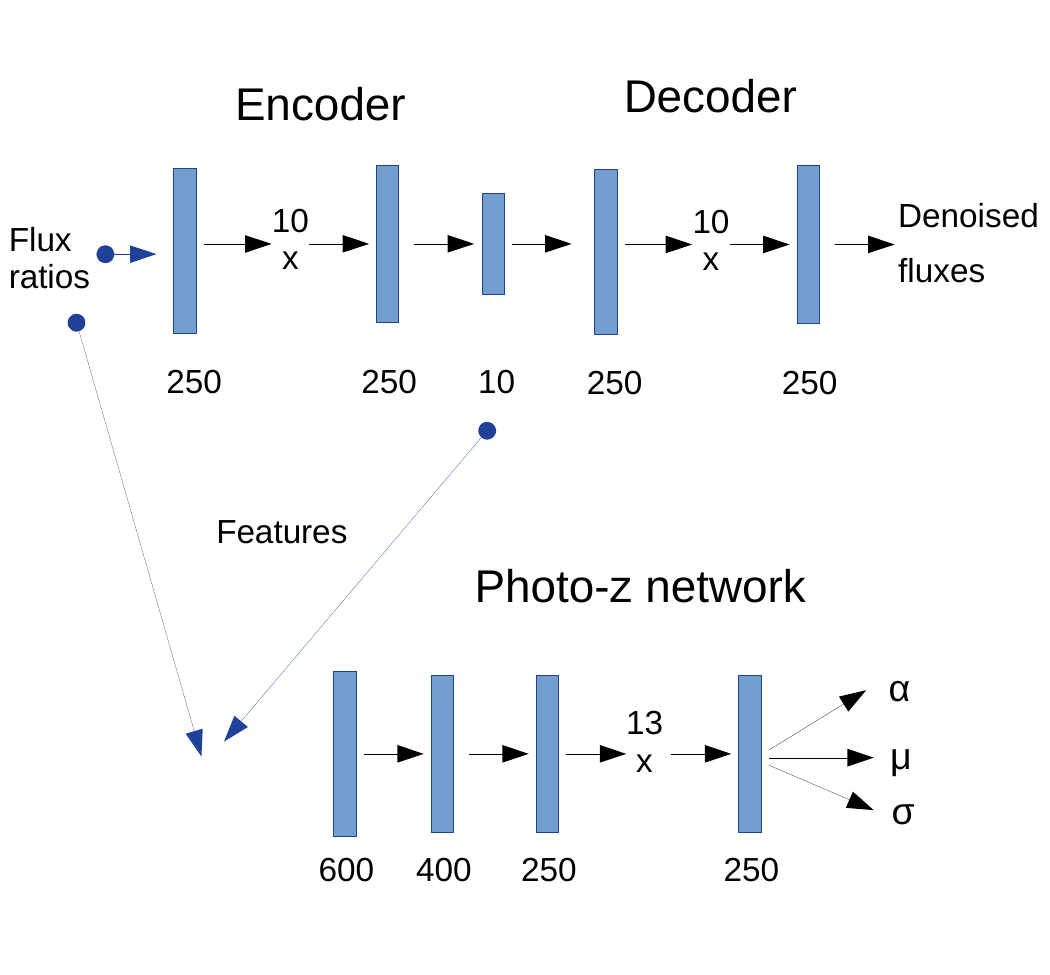}
\caption{The network architecture. \emph{Top:} The autoencoder, formed
by an encoder and a decoder network. The layers are linear and the figure
indicates the output dimension. Both networks include 10 layers with 
250 nodes. Following the intermediate linear layers are ReLU non-linearities,
batchnorm layer and a 2 per cent dropout after all linear layers, except the last three. \emph{Bottom:} 
We feed the galaxy flux ratios and the autoencoder
features into the photo-$z$
network. Here the
layers follow the same structure as the autoencoder, but with 1 per 
cent dropout after all linear layers. This network is a mixture density network and describe the redshift distribution as a linear mixture of 10 normal distributions.}
\label{fig_arch}
\end{figure}

Figure \ref{fig_arch} shows the network architecture of
\textsc{Deepz}, which uses a configuration with three linear
neural networks. The first two constitute an ‘autoencoder’: a
type of unsupervised neural network whose intent is to reduce
noise and extract features without knowing the redshift, making
it possible to train it with a larger dataset. We input the
flux ratios by dividing on the $i$-band flux. In the first step,
the ‘encoder’ maps raw information into a lower dimensionality
feature space, whereas the second step attempts to map it to the
original input data in the original dimensions.
The usage of the autoencoder is further discussed in \S \ref{auto_encoders}.

The network for predicting the photometric redshifts receives both the encoded
latent variables and the original input flux ratios. While the latent variables
include important information about the galaxy, this information alone
is insufficient for producing high precision PAUS redshifts. As discussed 
in \S \ref{encoder_on_fsps}, this is potentially due to the autoencoder not being
optimal for extracting sharp features in the spectra, like the emission
lines. The two sources of information are concatenated together before 
given to the network. Combining information processed in slightly different
ways is a common technique in machine learning (see e.g. \citep[][]{Huang2016}).

All three networks use linear layers. Each linear layer is followed by 
a batch normalization layer \citep{Ioffe2015} and a non-linear ReLU activation
function \citep{Nair2010}. In addition, we add dropout in selected places (see Fig. \ref{fig_arch} caption)
\citep{Nitish2014}. Instead of using linear layers, we have tested including 
a convolutional neural network (CNN) \citep{LeCun2004, Krizhevsky2017} for 
the PAUS fluxes. After testing various architectures, we 
conclude that adding a CNN component 
both degrades the photo-$z$ result and leads to a slower convergence. We therefore
use linear networks by default. The \textsc{Deepz}
predicts the galaxy redshift probability density
functions with the method described in the next
subsection.

\subsection{Predicting the probability density functions}
\label{pred_prob_func}
Estimating only the best fit redshift is insufficient for many science
applications (e.g. \citealt{Hoyle2018}). Often the users expect the photo-$z$ code to return a full
probability distribution, specifying how probable the galaxy actually is at
different redshifts. For a machine learning code, one might achieve this in different ways. The most straightforward approach is to bin the redshift range into classes and cast the problem into
a classification problem (e.g. \citealt{Gerdes2010}). In this way, the network can return a list of probabilities, 
each giving the probability of finding the galaxy in a given bin.

Alternatively, one can use a mixture density network (MDN) \citep{Bishop1994}.
In a MDN, the network outputs three vectors ($\mathbf{\beta}$,
$\mathbf{\mu}$ and $\mathbf{\sigma}$)  
that parametrise the probability distribution as follows:

\begin{equation}
p(z) \propto \sum_{i=1}^M \beta_i N(\mu_i, \sigma_i)
\end{equation}

\noindent
where  $N(\mu, \sigma)$ is a Gaussian with mean $\mu$ and standard 
deviation $\sigma$. The amplitudes ($\mathbf{\beta}$)
 give the relative contributions from each of the $M$ Gaussian components and sum to unity.
In this paper we use $M=10$, which is complex enough
 to capture the photo-z PDFs expected from simulations.
 This formalism can be adapted to use more general functions,
e.g. skewed Gaussian and Cauchy distributions. For simplicity we have restricted
ourselves to a linear combination of Gaussians, since this is a good
approximation for our data (\S \ref{pz_validate}). For the
redshift point-estimate value we use the mode (peak) of
the redshift probability density function (PDF).

Training the network requires a loss-function, which is the quantity that
one attempts to minimise. For training the MDN, we use the loss function

\begin{equation}
\text{loss} = - \sum_{\rm i} \log\left(p(z_{\rm label}^{\rm i})\right)
\label{loss_function}
\end{equation}

\noindent
where $z_{\rm label}$ is the redshift label (true redshift) and the sum is over a random subset of training galaxies (batch, see appendix \ref{deep_basics}). For observational data the label corresponds to  the spectroscopic redshift, while it is the true redshift in simulations. Minimising this expression is the same as maximising the probability. By default we predict the redshift PDFs using a MDN, but have
also tested the classification approach and will later comment on the differences.

\subsection{Training procedure} 
\label{training_procedure}
The network is trained on a graphical processor unit (GPU), using the loss
function (Eq. \ref{loss_function}) described in the previous subsection. We
minimise using a batch size of 100, meaning the gradients are computed using
100 galaxies. For the training procedure, we use the  Adam optimiser \citep{Kingma2014},
using a stepwise decaying learning rate. First we train 100 epochs with a learning rate $10^{-3}$ and then 200 epochs with
learning rates $10^{-4}$, $10^{-5}$ and $10^{-6}$, respectively in a 
decreasing manner. The network is first trained on simulations, which will
be presented in \S \ref{gal_simulations}, before optimising all weights in the network further 
with data. This simple approach works
well and is our default configuration. 

When pre-training on simulations, it is
critical to include noise. By default we add Gaussian noise with ${\rm SNR}=10$ (10\% error) and 35 (2.9\% error) for the narrow and broad bands, respectively. These values correspond to typical values for bright
galaxies observed with PAUS.
For future
work, we plan adjust the noise properties to closer
mimic the observed data.
Without adding noise to the simulations, the network
worked remarkably well on simulations, but could not adapt to the observed data. 
One can understand this from the features used by the network.
Without noise the
network can focus on some simple features, but it needs to use a combination
of them when the noise is introduced.

By default, the training is done with an 80-20 split, meaning 80 
and 20 percent of the sample are used for training and testing,
respectively. To generate the photo-$z$s for the full catalogue, the
network is trained 5 independent times so the training and test 
set never overlap. All figures use the same random split.

To avoid over-fitting hyper parameters, one should normally perform 
all optimisations on a separate validation set. We did not 
implement this from the start, mostly due to a small sample size. 
To avoid overfitting, we created a different random splitting
(still 80-20) before redoing the figures for the paper. We also
avoided overly finetuning e.g. the number of network layers. This
pragmatic solution avoids the most problematic cases of 
overfitting.

\section{Transfer learning from simulations}
\label{transfer_learning}
In \ref{transfer_learning_idea} we explain the concept of transfer learning,
while in \S \ref{gal_simulations} we describe the simulations, and 
\S \ref{results_pretraining} contains the main photo-$z$ results. Subsection \S
\ref{empty_bins} details the implications in  redshift ranges with fewer galaxies.

\subsection{Transfer learning}
\label{transfer_learning_idea}
Transfer learning is a common way of dealing with limited training sets \citep{Pan2010}.
Instead of training the model from scratch, one starts with a model that is
already trained on a different data set. This dataset is not required to look
identical to the dataset that one is interested in \citep{yosinski2014}. For example, the ImageNet 
curated image set with
millions of images and  associated classes is a common starting
point for training image classifiers \citep{Deng2009}. 
Using it as precursor training set leads to improved results and requiring less training.

The transfer learning approach often works by taking the network already trained for some
purpose. One then replaces the last layers (head) of the network, before
training the network on the data of interest. Often, this training focuses on training only the head of the network. This works since for image inputs the first layers of the network
pick up simple shapes, like strokes and edges. The features become
progressively more complex with the layers.

Transfer learning can work even when training on quite different data than
the domain of interest. This technique has successfully
been used for problems in e.g. supernova classification \citep{Vilalta2018}, data mining \citep{Schmidt2020} and 
Inertial Confinement Fusion (ICF) experiments \citep{Humbird2018}. In this 
paper we investigate the use of simulated galaxies to improve the photo-$z$ estimation. 
The generation of simulated galaxies has the advantage of providing an arbitrarily large training set, limited by the fidelity of the simulation.
This gap between observed data and simulations is expected to decrease as our
understanding of the PAUS data and simulations increases.

\subsection{Galaxy simulations}
\label{gal_simulations}
This paper investigates pretraining with two sets of simulations.
In \S \ref{template_sim} we present a template based simulation developed
for PAUS with realistic distributions in redshift, colour and 
galaxy properties to validate codes, estimate errors and compare 
with data. Then in \S \ref{fsps_sim} we describe the \textsc{FSPS}
simulation with a more sophisticated SED modelling. By default
this paper uses the \text{FSPS} simulation.

\subsubsection{Template based simulations}
\label{template_sim}
The magnitudes in this simulation are computed from the SED templates
taking into account the emission lines which are assigned following the recipes
described in Castander et. al (in prep.) and briefly described below.
First, we generate the rest-frame $r$-band luminosity applying an abundance
matching technique between the halo mass function and the Sloan Digital Sky
Survey (SDSS) luminosity function \citep{Blanton2003, Blanton2005}. Then, the galaxies
are evolved following evolutionary population synthesis models to their redshift.
Later, an SED and extinction are assigned to each galaxy by matching them
to the COSMOS catalogue of \citet{Ilbert2009} based on their
luminosity, colour and redshift. This means that the templates and extinction laws in
this simulation correspond to what is used in the COSMOS catalogue  of
\citet{Laigle2016}. From the ultra-violet (UV) flux, we compute the star formation rate,
and the flux of the ${\rm H}_\alpha$ line following \citet{Kennicutt1998}. This recipe is 
further adjusted to match the models of \citet{Pozzetti2016}.
The other line fluxes are computed following observed relations.
The SED, including the emission lines, is finally convolved
with the filter transmission curves to produce the broad and narrow-band fluxes.

\subsubsection{\textsc{FSPS} simulations}
\label{fsps_sim}
The main simulation in this paper is based on the 
Flexible Stellar Population Synthesis
(\textsc{FSPS}) code
\citep{fsps1, fsps2}. The \textsc{FSPS} code provides a state-of-the-art stellar population model and 
also a Python 
Application Programming Interface
(API)\footnote{http://dfm.io/python-fsps/}. We have extended the \textsc{FSPS} code to include
the PAUS filter transmissions.

\begin{table}
\begin{center}
\begin{tabular}{lll}
\toprule
Parameter & Range & Unit \\
\midrule
zred & [0, 1.2] & Redshift \\
logzsol & [-0.5, 0.2] & $Z / Z_{\odot}$ \\
tage & [0, 14] & Gyr \\
tau ($\tau$) & [0.1, 12] & Gyr \\
const ($k$) & [0, 0.25] & Fraction \\
sf\_start ($t_i$) & [0, 14] & Gyr \\
dust2 ($E(B-V)$) & [0, 0.6] & Colour \\
log\_gasu & [-4, 1] & Dimensionless \\
\bottomrule
\end{tabular}
\end{center}
\caption{The parameter ranges used in the simulations. The first column give the
FSPS-Python parameter name, with a corresponding symbol in parenthesis. The
simulations are generated by uniformly sampling within the ranges specified
in the second column. A third column state the parameter unit.}
\label{simulations}
\end{table}

Galaxies consist of a mixture of stars and dust. Stellar population synthesis
(SPS) models use the evolution of stars to model the galaxy properties. We refer
the reader to the \textsc{FSPS} papers for a description of the SPS formalism 
and only report briefly on our choice for various components. The star 
formation history (SFH) is an exponential decay model

\begin{equation}
\text{SFR}(t-t_i) = A \exp \left(-\tau (t - t_i) \right) + k
\end{equation}

\noindent
where $t_i$ parametrises the star-formation start for the galaxy and $\tau$ the exponential
decay. We have also included a component ($k$) with constant star formation. 
This choice of parameterization is known to fail to match the behaviour of late-type blue
galaxies and passive "red and dead" galaxies \citep{Vimal2014}. Using a
non-parametric SFH is a potential improvement to be considered in future
work. We note, however, that the simulations do not have to be perfect to benefit
from transfer learning (see \citealt{Pan2010}).

The stellar initial mass function (IMF) uses the \citet{Chabrier2003} model, 
while included nebular continuum and emission lines are from the FSPS 
integration with the \textsc{Cloudy} code \citep{Ferland2013, Byler2017}.
When producing the galaxy SEDs, the "age" parameter is fixed to the age at the 
redshift, using a Planck2015 cosmology \citep{Planck2015}. For dust extinction, we use 
the Calzetti extinction law (dust\_type=2, \citealt{Calzetti2000}),
parametrised by $E(B-V)$. When running, 
we set the metallicity of the gas equal to the metallicity of the galaxy, which the Python-FSPS document suggests. 
The emission lines are also  parametrised using a dimensionless 
gas ionisation fraction (log\_gasu), which is proportional to the 
flux of hydrogen ionising photons (Eq.1 in \citealt{Ferland2013}).

Table \ref{simulations} gives an overview of the parameter ranges used 
to generate the FSPS simulations. The simulations are generated by 
sampling each parameter uniformly within the given ranges and uncorrelated between the parameters. This parameter
distribution is obviously not realistic both in term of galaxy properties and colours. Note that we are only using the simulation to pre-train
the network, where we are less sensitive to the distribution exactly
weighting different galaxy properties.

\subsection{Photo-z with pre-training}
\label{results_pretraining}
\begin{figure}
\xfigure{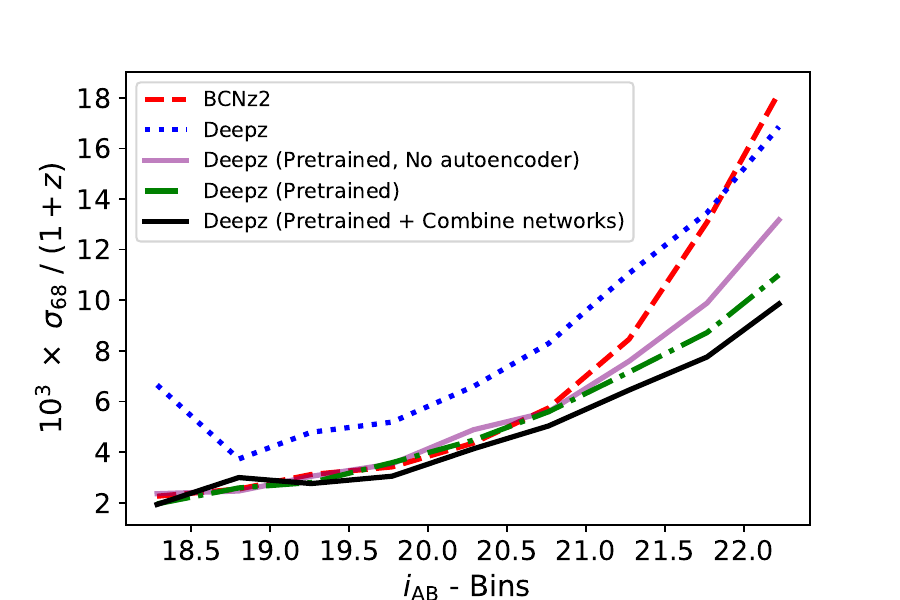}
\caption{
The $\sigma_{68} / (1+z)$ metric for 100\% of the galaxies with secure redshifts in magnitude bins and for different codes. The dashed (red) line is the baseline performance and it corresponds to the 
\textsc{BCNz2} results from \citet{Eriksen2019}. The rest of the lines show the results for various 
\textsc{DEEPz} configurations.}
\label{main_photoz}
\end{figure}

Figure \ref{main_photoz} shows the main photo-$z$ results, which uses
the training procedure explained in \S \ref{data_architecture}. For quantifying
the photo-$z$ performance, we define

\begin{equation}
\sigma_{68} \equiv 0.5\left(z_{\rm quant}^{84.1} - z_{\rm quant}^{15.9}\right)
\end{equation}

\noindent
which is half the difference between the 84.1 and 15.9 percentile. The 
$\sigma_{68}$ corresponds to the standard deviation for a Gaussian
distribution, but  is less sensitive to outliers. Throughout the paper 
we also use a strict outlier defined by

\begin{equation}
|z_{\rm p} - z_{\rm s}|\ /\ (1 + z_{\rm s}) > 0.02    
\label{outlier_def}
\end{equation}

\noindent
where $z_{\rm p}$ and $z_{\rm s}$ are the photometric and spectroscopic redshift,
respectively. We label this outlier
fraction "strict", since it should not be confused with what is
an outlier in a broad-band survey. In a broad-band survey
the photo-$z$ scatter is much larger and the corresponding outlier
definition (Eq. \ref{outlier_def}) is often 10 times more relaxed 
\citep{Kuijken2015, Bilicki2018}.

The dashed line (Fig. \ref{main_photoz}) shows the photo-$z$ scatter using
the \textsc{BCNz2} template fitting code as a function of differential
$i$-band
values. The dotted line shows the performance when training \textsc{Deepz} only on observed data.
The photo-$z$ scatter is significantly larger than for \textsc{BCNz2},
except for the faintest magnitudes ($21.8 < i_{\rm AB}$). 
Pre-training the network on simulations before training with data
reduces the photo-$z$ scatter by ~50\% at the faint end. Not including an autoencoder,
 as discussed further in the next section (\S \ref{auto_encoders}), degrades the performance at the faint end. Lastly, the solid line shows the result when training the networks 10 different times
with multiple networks (see \S \ref{training_procedure} and \S \ref{multiple_predictions}). These are the 
currently best \textsc{Deepz} results. In appendix \ref{photoz_scatter} we have included a photo-$z$ versus spec-$z$ plot to highlight the outliers.

When pretraining with either the \textsc{FSPS} and template
simulations (\S \ref{gal_simulations}), we find a significant
reduction in the photo-$z$ scatter. For the cases of no-pretraining, pretraining
on template simulations and pretraining on the \textsc{FSPS}
simulations, the $\sigma_{68} / (1+z)$ without quality cuts
is 0.0095, 0.0077 and 0.0069,
respectively. This indicates that a better SED 
modelling is more important than a correct colour space 
distribution for simulation used for pre-training. We have also tested generating the \textsc{FSPS} simulations 
fixing the gas ionisation fraction, which gave a slightly 
higher scatter. Other approaches to improve the simulations 
could lead to an even better performance.

\subsection{Redshift intervals without spectroscopic galaxies}
\label{empty_bins}
A fundamental limitation when training the PAUS photo-$z$ is the small training set. Deep neural networks are often trained with millions of training samples, e.g.
in ImageNet \citep{Deng2009}. Transfer learning from simulation is one approach
for reducing the required number of spectroscopic galaxies.

\begin{figure}
\begin{center}
\includegraphics[width=0.49\textwidth]{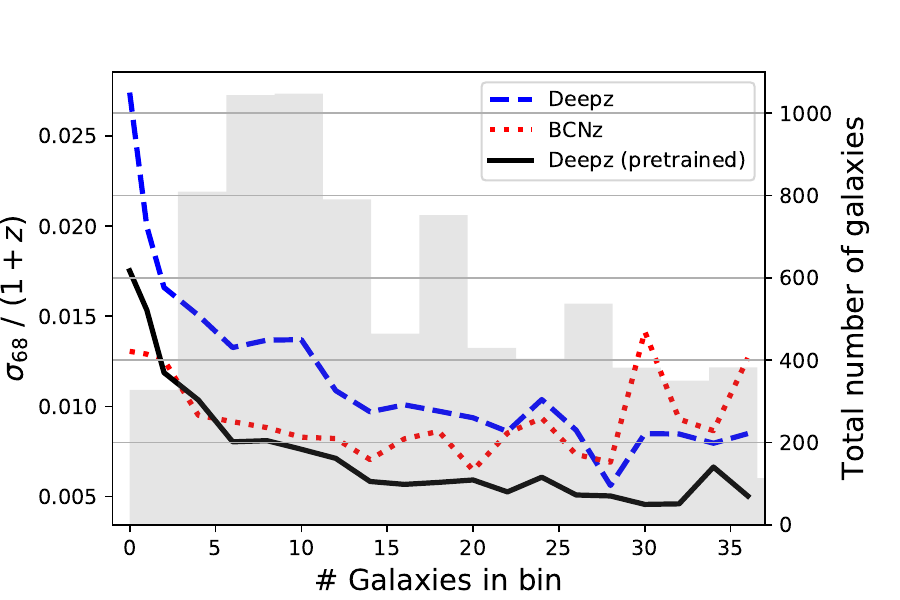}
\end{center}
\caption{The effect of redshift ranges with a smaller number of galaxies. On the x-axis is the number of galaxies in bins of
$\Delta z = 0.001$. The  dotted 
line shows the \textsc{BCNz2} result, while the continues and dotted lines show the \textsc{Deepz} when pretraining
or not on simulations. The shaded histogram displays the total number of galaxies for each
value on the x-axis.}
\label{missing_zbins}
\end{figure}

Figure \ref{missing_zbins} shows the photo-$z$ scatter as a function of the number of galaxies in the bin for bins of 
 $\Delta z = 0.001$. We want to understand how the density of 
 spectroscopic redshifts affects the photo-z scatter.
These bins are only used to illustrate the effect of the density
and are not used when training the MDN. With the \text{Deepz} code, the photo-$z$
scatter is clearly higher in bins with only a few galaxies. A dotted line 
shows the \textsc{BCNz2} result which is much less affected
by the number of galaxies per bin, specially for very sparse bins.
This shows that the number of galaxies in the bin is the underlying reason 
and not by bins with few galaxies indirectly select higher redshifts.
Pretraining on simulations reduces the difference, but there is still
a region with fewer galaxies where the template fitting works better.
Lastly, appendix \ref{label_smoothing} details how to deal with low
density regions for networks without a MDN.

In addition, we have tested using the mixup \citep{Zhang2017}
method of data augmentation. Normally data augmentation requires knowing which
transformation can be applied without changing the meaning of the
data. For example, when classifying images one might want to include 
rotations and slightly changing the brightness. Instead, the
mixup method uses the linear combination of a random pair of inputs.
Applying this technique to our data did not improve the photo-$z$ 
scatter.

\section{Autoencoders}
\label{auto_encoders}
The network architecture includes an autoencoder (see \S \ref{network_architecture}). 
Section \ref{auto_encoders_idea} explains with a single SED example how autoencoders 
can reduce the observational noise and extract features. Then in \S \ref{encoder_on_fsps} we discuss application of the technique to
our \textsc{FSPS} simulations and discusses the impact for redshift estimates.

\subsection{Autoencoders}
\label{auto_encoders_idea}

Figure \ref{fig_arch} (top) of the \textsc{Deepz} network architecture 
shows the two autoencoder networks. The encoder network transforms its 
input into the latent or feature space. In our case, the
input is 46 bands (40 NB, 6 BB) and the latent space has 10
variables, which is a reasonable number of parameters to describe a
galaxy SED. A decoder network then attempts to reconstruct the input. One
can train these networks with a loss function comparing the recovered 
values and the original input. Since the latent space is smaller than 
the input, the autoencoder is required to compress the information. 
The noise can not be compressed to fewer numbers and therefore
gets removed.

\begin{figure}
\xfigure{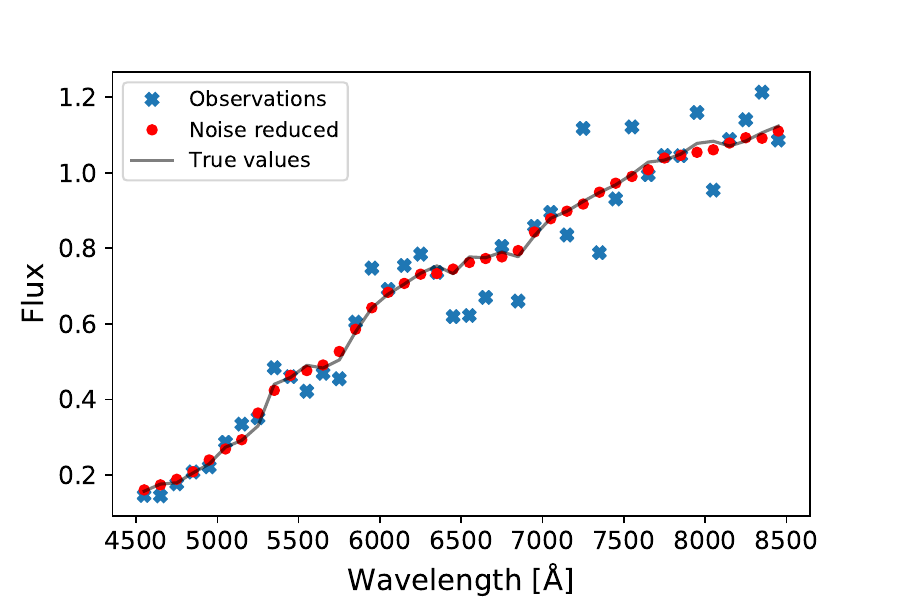}
\caption{Effect of the denoising network for one
example galaxy. The simulation is generated from a single elliptical SED with arbitrary
flux units and     
a uniform redshift distribution.
The crosses and circles show the input and denoised
narrow-band measurements, respectively. A solid line displays the noiseless flux
of the SED.}
\label{noise_red}
\end{figure}

To illustrate how the autoencoder works, we have generated a set of simple
simulations. Using a single elliptical SED (Ell1\_A\_0) that was 
used both in the COSMOS2015 \citep{Laigle2016} and the PAUS photo-$z$ papers 
\citep{Eriksen2019}, we estimate galaxy fluxes for a uniform redshift
distribution. We added Gaussian noise with $\text{SNR}=10$ 
and 35 for the narrow and broad bands, respectively, which corresponds to
the noise level for a
bright PAUS galaxy. This simulation
is then used to train an autoencoder. Figure \ref{noise_red} compares 
the input, true and noise reduced fluxes for a typical case. The recovered 
output has clearly reduced noise. The autoencoder achieves this by using the fact that galaxies
in this simulation do not populate the full colour space, but a 2D
sub-manifold described by the redshift and amplitude.

Note that an autoencoder  can also be applied to broad bands alone, where the input dimension is typically smaller than the latent space. With the
method above, the autoencoder would simply become the identity
mapping. This can be solved by adding Gaussian noise to the input
fluxes \citep{Vincent2010}.

\subsection{Tests on \textsc{FSPS} simulations}
\label{encoder_on_fsps}
\begin{figure}
\xfigure{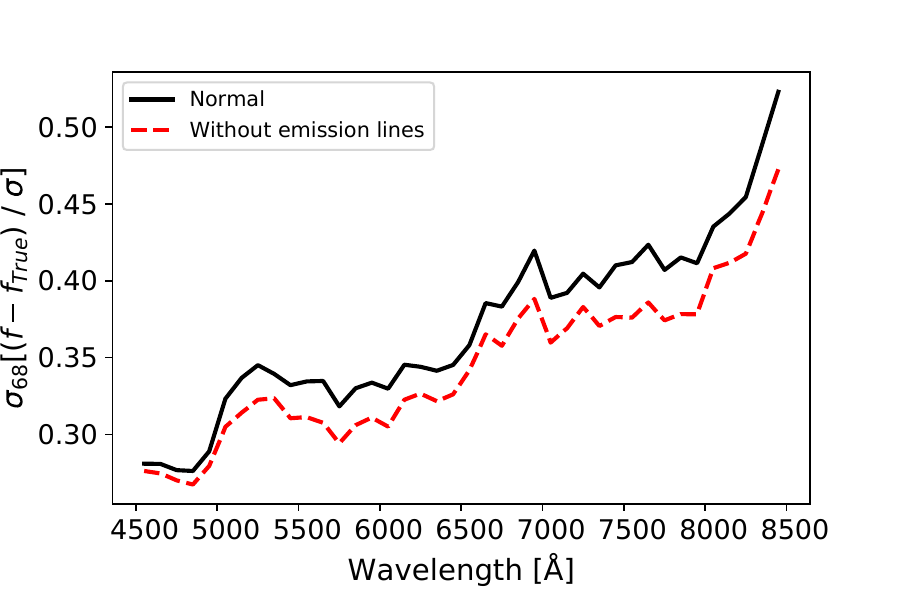}
\includegraphics[width=0.475\textwidth]{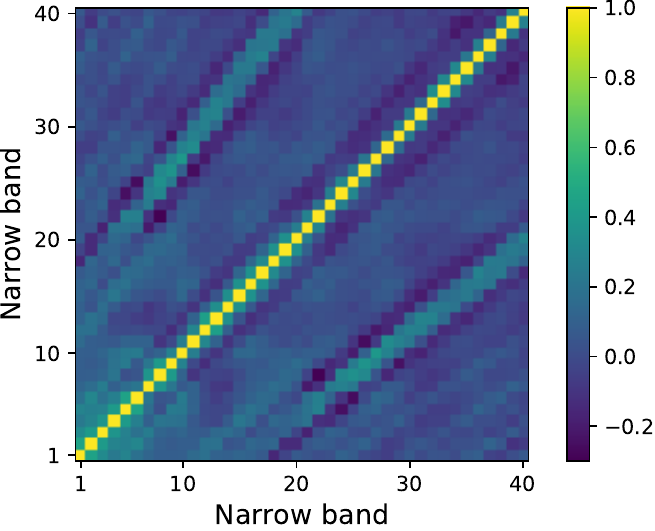}
\caption{\emph{Top:} The scatter ($ \sigma_{68}$) of the difference of the difference of the 
denoised ($f$) and true fluxes ($f_{\rm True}$) relative the known errors of the input fluxes ($\sigma$). In the dashed line the bands with emission lines are removed. \emph{Bottom:}
The correlation matrix of the denoised flux between different narrow bands.}
\label{autoencoder}
\end{figure}

Figure \ref{autoencoder} quantifies the impact of using an autoencoder on
the 
\textsc{FSPS} simulations (\S \ref{gal_simulations}). The top panel compares 
the error in the recovered fluxes with the input error, as 
a function of wavelength. A unity mapping would give a horizontal line at unity.
When using the autoencoder, we find the flux errors decrease. For 
the blue bands the error is 30\% of the expected value and it increases
to 50\% for the redder bands. For the broad bands, the ratio between 
the recovered and input error is 1.04, 0.72, 0.66, 0.61, 0.22 and 
0.97 for the $uBVriz$ bands, respectively. A problem is that the
autoencoder
smooths the emission lines (see dashed line), which is a known artefact in autoencoders 
\citep{Dosovitskiy2016}. The recovered fluxes are good for training the
redshift network, but should be used with caution for other 
scientific applications, e.g. estimating the mean flux.

The bottom panel (Fig. \ref{autoencoder}) shows the correlation between the
different narrow bands. Here the broad bands are used to train the network,
but not included in
the figure for clarity. When using an autoencoder, the galaxy is transformed
by the encoder into the latent space variables, which
describe the galaxy. This transformation is affected by noise in the
input and is also not perfect and this introduces an error on the latent
variables. When reconstructing the fluxes with the decoder, this creates correlated noise between different bands. This can be understood from
the latent space representing information related to galaxy type or
dust properties. As can be seen, the correlation is strongest with nearby 
bands. Furthermore, there is a correlation between bands that are separated by 
1500\AA, resulting from confusing the $\text{O}_{\text{II}}$ and 
$\text{O}_\text{III}$ lines. 

When training the redshift network (Fig. \ref{fig_arch}), we combine the
information from the input fluxes and features produced by the
autoencoder. Combining information processed in different ways
together is a standard technique in deep learning (see e.g. \citealt{Huang2016}). 
When training on the simulations, we combine the loss from both the 
autoencoder and the redshift estimation, while ignoring the autoencoder 
loss when fine-tuning on data. Fig.
\ref{main_photoz} includes a line where the
auto-encoder is disabled by setting all
features to zero. This shows the autoencoder
has a significant impact on the photo-$z$ scatter for faint galaxies. We also expect the autoencoders to become more 
important when training the autoencoder with data in the wide fields
(CFHTLenS W1 and W3) without spectra. We leave this for future work.

\section{Adding information from individual exposures}
\label{individual_exposures}

We describe in \S \ref{ind_exp_idea}  the motivation of including information from individual exposures
when training the network, while \S \ref{multiple_predictions} 
explores combining multiple networks to reduce the errors. Lastly,
\S \ref{ind_exp_test} studies the use of individual exposures at 
test time.

\subsection{Incorporating individual exposures}
\label{ind_exp_idea}
Astronomical surveys perform repeated measurements over the same parts of the
sky in systematic patterns. The purpose of making multiple observations is
often to produce a combined measurement with reduced noise, allowing
the observation of fainter objects. For example, the Dark Energy
Survey (DES) \citep{Hoyle2018} and the Kilo-Degree Survey (KIDS, \citealt{Kuijken2019})
have imaged each position $\sim 8$, $4-5$ times in each band, respectively. The 
Rubin Observatory Legacy Survey of Space and Time (LSST) will measure each location several hundred times \citep{LSST2009}. In PAUS, the  COSMOS field is nominally imaged at least 5 times in each narrow band.   

For estimating the redshifts, the individual measurements are typically  first combined into coadded fluxes. A standard choice is to combine the individual measurements by an inverse variance  weighting, which is statistically optimal for a combination of independent Gaussian measurements. However, this combination is not optimal if there are photometric outliers. These outliers can arise from  multiple sources   including scattered light \citep{Cabayol2019}, electronic cross-talk between  the charge-coupled devices (CCDs) or data reduction issues in the calibration or photometry.  

Removing problematic measurements is difficult. The PAU data management (PAUdm)  code flags many of the problematic outliers based on image diagnostics. Outliers are however still present in the PAUS data. The PAUS observations are often noisy (${\rm SNR} < 1$) and for many (galaxy, band) combinations, we only have 3 exposures after flagging measurements, making the detection of  outliers for a single band hard. Some outliers, like those resulting from negative cross-talk, are clearly visible, since the flux is much lower than nearby bands. However, positive flux outliers are harder to flag and are problematic since they can be confused with emission lines, leading  to photo-$z$ outliers.  

Instead of manually removing measurements, we want the photo-$z$ code to select itself the correct measurements by working directly with the individual exposures.
 The most obvious approach would be
to directly input the individual exposures to the network. However, multiple problems
arise when applying the technique to observational data. For example, PAUS 
has a minimum of 5 exposures in the COSMOS field, however, many of the
observations are removed since they contain bad data. Also, there are
regions with more than 5 exposures. This means the input to the photo-$z$
code would not be a dense array with all values present.

Furthermore, inputting all measurements individually drastically increases the
network size. In addition to at least increasing the network with 5
times the inputs (number of exposures), one should also inform the network
which measurements are present. If specifying a mask, this would lead to
another doubling of the input. Also, the ordering of  individual
fluxes is not unique. Appendix \ref{all_permutation} details how this
problem can partially be overcome by permuting the order of the 
individual flux measurements when training. This approach does not solve 
the issue to the required accuracy.

An alternative approach builds on the technique of data augmentation. When training neural networks, it is common to perturb the input to produce a
slightly different input. For example, one might crop, flip or adjust the
colours of an image. This produces images humans essentially see as unchanged, but
appear different to the network. Adding these permutations often ends
up improving the performance and is standard for many applications \citep{Perez2017}.

The approach suggested in this paper is training the network with randomised
coadds, constructed on the fly with a randomised selection of individual exposures.
Each time when training the network with a set
of galaxies, the individual exposures are chosen to be included with a 
probability $\alpha$. Since the coadded fluxes are constructed for each epoch, 
it means each galaxy will look different to
the network at each epoch. We have tested two methods to handle galaxies not having measurements in all bands after the random selection. In one, the galaxy is removed for
a specific epoch when the randomisation leads to not having measurements in
all bands and the second modified the sampling method to ensure at least one measurement
is present in each band. The
construction of randomised coadds can be computed simultaneously on a GPU without
significant computational overhead\footnote{
The coadded fluxes are generated on the GPU
by inputting the individual fluxes in a dense matrix. A Bernoulli distribution
with fixed probabilities is used to determine if a measurement should be
included or not. We then generate the included from the include exposure by an inverse
variance weighting. In benchmarks on an NVIDIA Titan-V, this operation only
adds 0.02ms for 1000 galaxies.}.

\begin{figure}
\xfigure{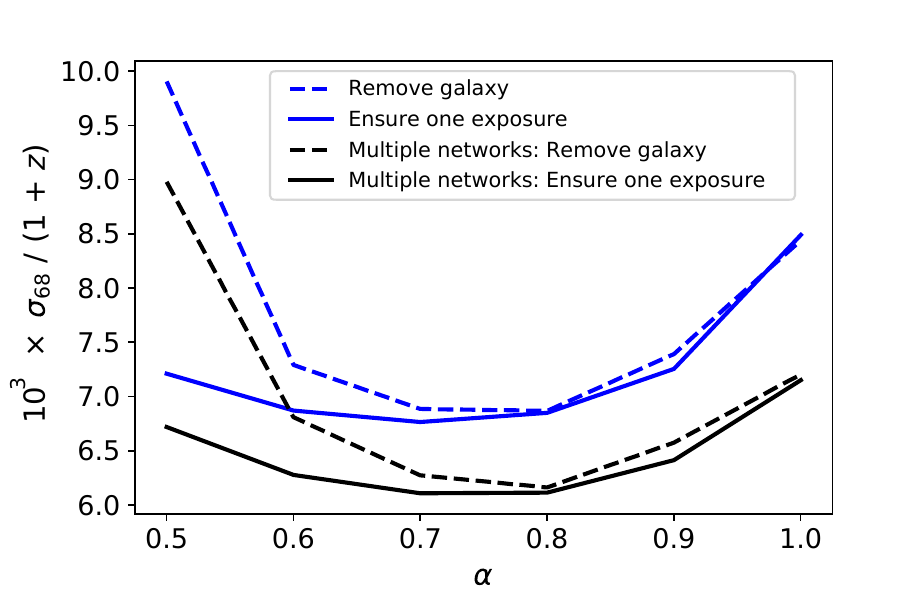}
\caption{The $\sigma_{68} / (1+z)$ when varying the value of the probability of including an exposure in the coadd when
training ($\alpha$). A factor of $\alpha=1$ corresponds to no 
randomness. Blue lines
use a single network, while black lines combine multiple networks. For the
dashed 
lines a galaxy is removed when the randomness leads to a galaxy not having
measurement in all bands. Continuous lines use a randomisation procedure which is
required to keep at least on measurement per band.}
\label{alpha}
\end{figure}

Figure \ref{alpha} shows the photo-$z$ scatter for different probabilities
of using the individual measurements ($\alpha$).
Including this randomness
when training significantly
reduces the photo-$z$ scatter. When predicting with a single network,
the photo-$z$ error decreases by 20\% compared with no randomisation. The dashed lines show the result when
one removes galaxies which do not have measurements in all 40 bands after
applying the random exposure removal. Note that which galaxies are removed
depends on the epoch, since the networks see each galaxy multiple times
(appendix \ref{deep_basics}). Below about $\alpha=0.7$, the networks performance degrades, which follows from many galaxies not being used in the training if
not ensuring one exposure being present. By default results in
this paper use $\alpha = 0.8$. The result for the "Multiple networks" will be 
discussed in \S \ref{multiple_predictions}.

Training a neural network means learning a mapping between the observed colours
and the redshifts. 
In this process, the network 
also needs
to discover which features are real or simply due to bad photometry. The
randomised construction of the coadds when training leads to the network
seeing the same galaxy with and without problematic measurements. This
makes it easier to learn which features are properties of the galaxies,
like the emission lines. This method is expected to be less effective in the limit of an
infinite training sample. However, the randomisation makes an important
difference for a limited training set
with outlier measurements.

\subsection{Combining predictions from multiple networks}
\label{multiple_predictions}

The photometric redshift results discussed until now have 
used a 80-20 split between the training and test sample 
(see \S \ref{training_procedure}). 

One could attempt to change the splitting ratio (e.g. 90-10)
to increase the
number of galaxies used for training. In the extreme limit one
would have one network per galaxy, which would be prohibitively
computationally expensive. Instead we focus on combining multiple networks
and have defined ten (random) ways of splitting the catalogue into a training 
and a test sample. With this approach, one can train and combine the PDFs
for multiple networks for each galaxy in the training set. Note, the estimated
photo-$z$ always use networks which have not been trained with the same galaxy.

Figure \ref{main_photoz}, which compares the effect from different ideas,
includes a line showing the photo-$z$ predictions using multiple networks. The 
photo-$z$ results shown correspond to training with ten different 80-20 splits and  
then averaging the resulting $p(z)$ distributions. This means training
the networks in total 50 times. Combining the networks leads to about 10
percent lower photo-$z$ scatter for the faintest galaxies in the sample
($i_{\rm AB} = 22.5$). We also tested generating the photo-$z$ using 100
different splits. The benefit of multiple networks saturated with fewer than
10 splits, which we use by default in the \textsc{Deepz} code.

In Figure \ref{alpha} we also study the effect of combining multiple
networks when randomly creating coadds. The 
two blue lines corresponding to a single network. Two black lines show the performance
combining multiple networks. The photo-$z$ scatter for the two methods
follows a similar trend. This result shows that combining multiple networks, 
rather than being redundant, is an improvement on top of the coadd 
randomization.

\subsection{Test-time augmentation}
\label{ind_exp_test}
In the previous subsection we applied data augmentation when training
the network. Data augmentation can also be used when inferring the redshift, 
often named test-time augmentation and can be applied in addition to the 
training augmentation discussed in the previous subsection.

Training a neural network is often computationally 
expensive, although for our case, the training is faster than the 
\textsc{BCNz2} template fitting method\footnote{
Training neural networks can be computational demanding, but
is accelerated with GPUs. Evaluating neural networks can be extremely fast. For
determining galaxy redshifts, the \textsc{BCNz2} algorithm ended up taking around
30 seconds per galaxy. In contrast, neural network algorithms with better
results determine the redshift of 12000 galaxies per seconds on a single 
Titan-V GPU. Ignoring the training time, this is a speedup of 360000
times.}.
Predicting the redshift is very fast with neural networks. This allows
studying how the photo-$z$ is affected by changes in the photometry. In
this section, we have tested systematically removing individual
fluxes, constructed the coadded fluxes and estimated the corresponding
photometric redshifts.

\begin{figure}
\xfigure{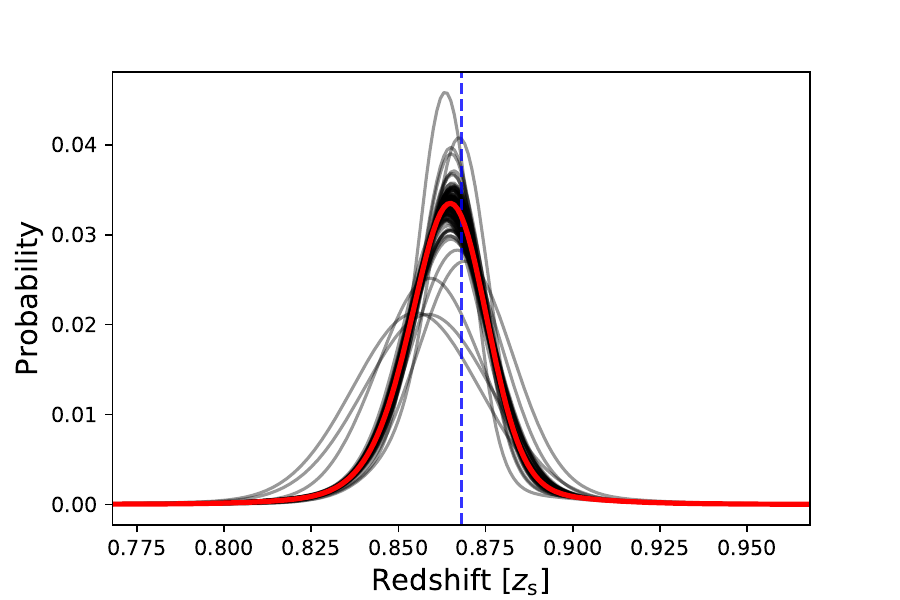}
\caption{Test time augmentation, removing individual flux measurements for a single
galaxy. The vertical line indicates the redshift, while the solid red line gives
the $p(z)$ using the full coadd. The thin lines show the $p(z)$ estimated without
individual flux measurements.}
\label{indexp_test_aug}
\end{figure}

Figure \ref{indexp_test_aug} shows the effect of dropping different exposures
for an example galaxy. Here the vertical line marks the spectroscopic (true) redshift, 
the thin lines show the $p(z)$ for different removed exposures and the 
solid red line shows the $p(z)$ estimated from the coadds. In most cases, the 
$p(z)$ distributions peak at a redshift that is slightly shifted from the 
spectroscopic redshift. When dropping one of the exposures, the $p(z)$ 
prediction peaks around the spectroscopic redshift. In other cases, dropping 
a single exposure leads to the $p(z)$ moving in the wrong direction and 
therefore produces an outlier. From this experiment, we conclude that
systematically estimating the photo-$z$ by dropping individual measurements 
is not a viable strategy.

\section{Validating the redshift distribution}
\label{validation}
In this section we validate whether the redshift probability
distributions accurately represent the
uncertainties (\S \ref{pz_validate}). We
also introduce redshift quality cuts (\S \ref{quality_cuts}) to select subsamples
with better redshift determination.

\subsection{Validating the redshift distributions}
\label{pz_validate}

The \textsc{Deepz} code does not only predict a point estimate, but also
the redshift probability density. Knowing the redshift distribution for
each object is useful for various applications, like e.g. weak 
gravitational lensing measurements. For this reason, it is important 
that the PDFs actually represent the redshift uncertainty, not simply
peaking around the correct redshift.

A common approach for testing the quality of the probability distribution
is the probability integral transform (PIT, \citealt{Dawid1984, Gneiting2005, Bordoloi2010})

\begin{equation}
\text{PIT} = \int_0^{zs} dz' p(z')
\end{equation}

\noindent
where $p(z)$ is the probability distribution and the integration is
from zero to the spectroscopic redshift ($z_s$). If the probability
distribution estimate actually represents the underlying distribution, the
distribution of PIT values would be uniform.

\begin{figure}
\xfigure{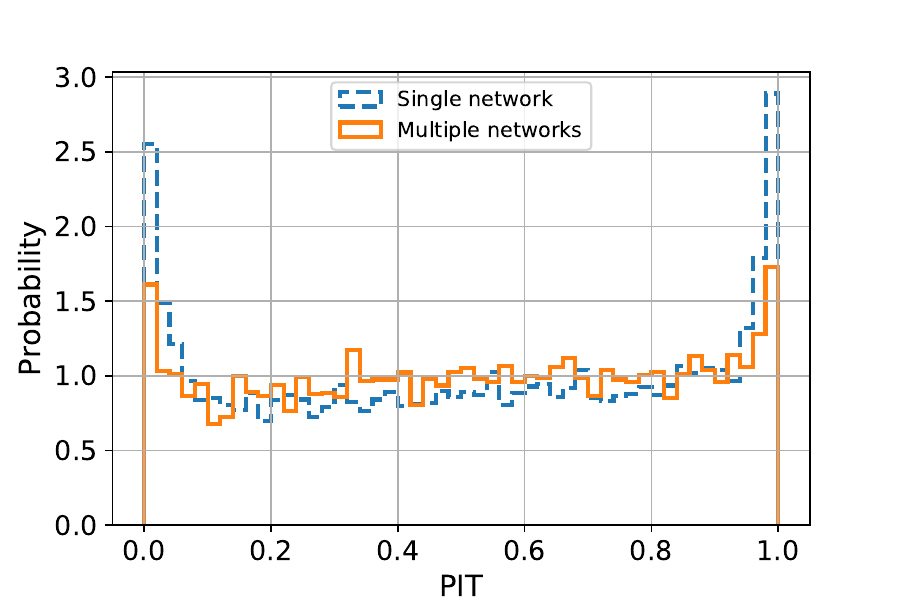}
\caption{Testing the $p(z)$ distributions using the PIT distribution. The solid line
shows the result when combining multiple networks, while the dashed line shows
the result for a single network.}
\label{pit_distribution}
\end{figure}

Figure \ref{pit_distribution} shows the PIT distribution for \textsc{Deepz} of the test set. The dashed
line shows the
the result for a single network, while the solid line shows the result for
multiple networks. The distributions are close to uniform, except
for low and high PIT values. These peaks correspond to photo-$z$ outliers which
are not reflected in the PDFs predicted by the network. The main contribution 
behind the drop when combining multiple network is the combined networks reduce 
the outlier rate, making the $p(z)$ simpler to estimate.

The uniformity of the PIT diagram should not be taken for granted. In
addition to problems with outliers, many redshift codes have a problem,
 underpredicting the width of the redshift PDFs \citep{Schmidt2020}.
In early versions of this work, we  predicted the probability distribution
using a classifier, binning the galaxies in different bins. The resulting
PIT histograms were not sufficiently flat. In \citet{Guo2017} 
the authors claim that the classical neural networks have PDFs that are relatively 
well calibrated, but this is no longer the case when
dealing with modern architectures. These include many components, like the batch
normalisation and weight decay, which leads to reported probabilities to not
accurately represent the true distribution. Using a mixture
density network (\S \ref{pred_prob_func}) provides better probability distributions for our application.

\subsection{Quality cuts}
\label{quality_cuts}

\begin{figure}
\xfigure{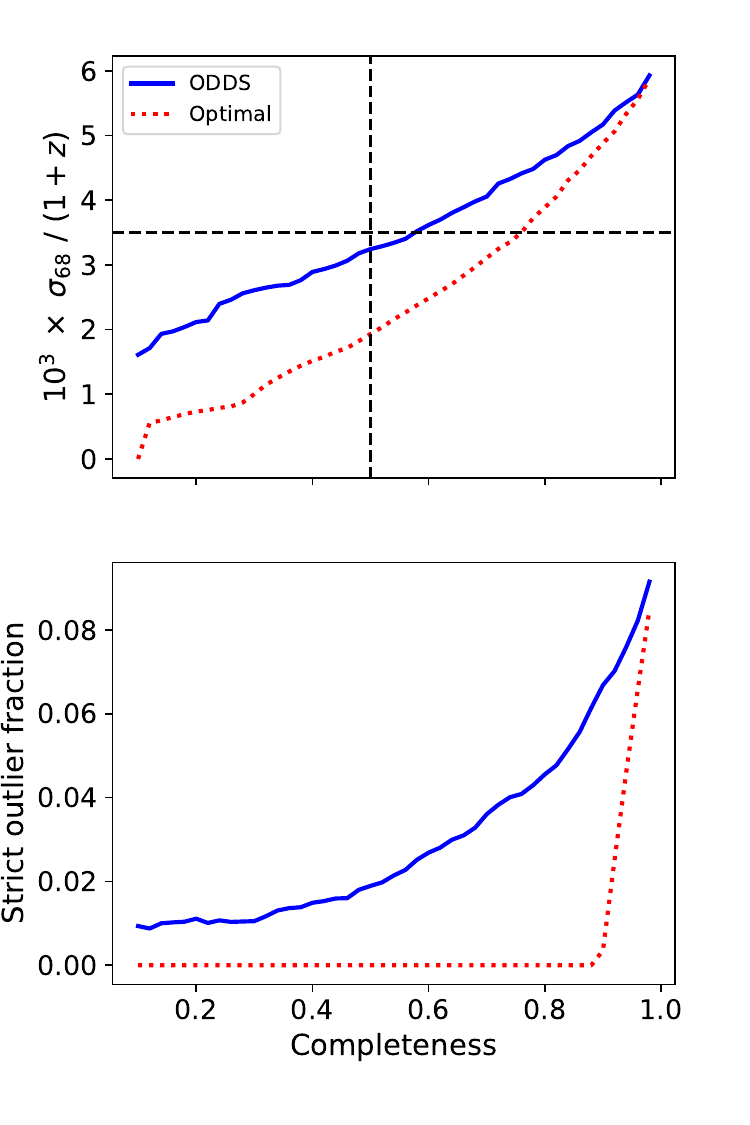}
\caption{The effect of introducing photo-$z$ quality cuts for the secure redshift
sample to $i_{\rm AB} < 22.5$. The top and bottom 
panels show the photo-$z$ scatter and outlier rate, respectively. Continuous lines cuts 
based on the ODDS parameter, defined from the probability distribution.
The optimal lines cut based on the spectroscopic redshifts to demonstrate
the (idealistic) lower limit of a quality cuts.
The horizontal line in the top panel corresponds to PAUS photo-z target for
a selected 50\% of the sample.
}
\label{fig_quality_cuts}
\end{figure}

This subsection studies introducing 
redshift quality cuts for the \textsc{Deepz} code, but one should be aware
of the potential side effects of these cuts.
For different science applications, one might want to select a subset of galaxies with
higher photometric redshift precision, e.g. to cross-correlate galaxy
counts with other samples to estimate the photo-$z$ scatter between redshift bins. A 
common problem with cutting on photometric redshift quality is 
unintentionally introducing clustering, since the quality might be tracing
spatial patterns like observing conditions \citep{Ross2011, Marti2014_cuts}.
In \citet{Eriksen2019} we reported on visible spatial patterns in the quality
of the \textsc{BCNz2} template fitting.
The ODDS quality parameter
introduced in \textsc{BPZ} \citep{Benitez2000} is defined by 

\newcommand{\zpeak}{z_{0}} 
\newcommand{\zdelta}{\Delta z}

\begin{equation}
\text{ODDS} \equiv \int_{\zpeak - \zdelta / 2}^{\zpeak + \zdelta / 2} dz\, p(z)
\end{equation}

\noindent
where $p(z)$ is the probability distribution, $\zpeak$ its mode and
$\zdelta = 0.003$ being the fixed interval around 
the most likely redshift (mode of the distribution).

Figure \ref{fig_quality_cuts} shows the photo-$z$ scatter (top) and strict
outlier rate defined in Eq. \ref{outlier_def} (bottom) as a function
of the completeness, which is
the fraction of galaxies kept after the cut. Introducing a quality
cut based on ODDS gives a significantly better photo-$z$ scatter and
outlier rate. The PAUS \textsc{Deepz} redshifts for 50\% now clearly surpass the target
performance of $\sigma_{68} = 0.0035(1+z)$ to $i_{\rm AB} = 22.5$. 
It is likely that the scatter is higher, since galaxy types lacking spectral coverage will probably have a lower quality photo-$z$ estimate. The optimal lines select by $|z_p - z_s|$ using the spectroscopic information
and indicate there might be further room for improving the quality cut.

In  \citet{Eriksen2019} we tested the performance for a set of
quality parameters. There we also used the pz\_width quality parameter that
measures the distance between the 1 and 99 percentile of the PDFs. For 
\textsc{Deepz}, we find that this quality parameter performs worse. 
By default, the \textsc{BCNz2} results were reported using an
adjusted version of the Qz parameter \citep{Brammer2008}, which is
a multiplicative combination of the ODDS, the pz\_width parameter
and the $\chi^2$ of the fit. Unlike a template fitting code, the MDN 
network of \textsc{Deepz} directly estimates the $p(z)$ with normalisation. Therefore we cannot use the same quality cut.

In \S \ref{individual_exposures} we introduced a technique of 
randomly generating the coadds when training the network. We have
tested generating the photo-$z$ for these different coadds based
on 80\% of the exposures and then estimating the variance between the different photo-$z$ estimates.
 Our initial expectation, was that smaller photo-$z$
variations would indicate a more secure photometric redshift determination. Actually, 
often the opposite is true. When there are very small variations when 
removing exposures, a subset of the exposures tends to drive the 
photo-$z$ solution. Cutting to keep galaxies with a higher variability in 
the predictions tends to perform better. However, this is a weaker quality 
cut than e.g. the ODDS.

\section{Conclusions}
\label{conclusions}
In this paper we introduced  a new deep learning photo-$z$ code,
\textsc{Deepz}. We uses the Physics of the
Accelerating Universe Survey (PAUS), which has 40 narrow bands \citep{Padilla2019}, as a test case.
Previous work 
showed how PAUS can achieve the target photo-$z$ precision using a 
template based fitting code \citep{Eriksen2019}. 
This in itself is a non-trivial result, since 
previous attempts to apply \textsc{ANNz} and \textsc{DNF}
\citep{devicente2016}
to PAUS simulations were unsuccessful. 
The standard \textsc{ANNz} essentially 
ignored the narrow bands because of their lower signal to
noise ratio.
Also, the lack of sufficient training data resulted in
the codes being unable to reach the target photo-$z$ precision. In this 
paper we introduced a machine learning approach to overcome this
obstacle and obtained state-of-the-art PAUS redshift precision.

The network was trained using flux ratios from the 40 
PAUS narrow bands, combined with the CFHTLenS $u$-band and $BVriz$ bands from the Subaru telescope in the COSMOS field. 
The network inputs are the 46 fluxes, normalised to the $i$-band. To train the network,
we used the zCOSMOS DR3 catalogue, limited to secure redshifts and simulations.
The network was implemented using the \textsc{PyTorch} \citep{Paszke2017} 
library, a widely used framework in the deep learning research community. Our
architecture consisted of three different networks, an autoencoder to 
extract information about the galaxy and a network to predict 
the redshift. The network estimated the full PDF using 
a mixture density network (MDN) \citep{Bishop1994} and the final distribution is the mean redshift PDF from an ensemble of
10 different networks.

The application of the machine learning approach based on only observed data 
as a training shows worse performance than the template method (\textsc{BCNz2}).
However, transfer learning from simulations improves the photo-$z$ precision, 
especially for faint magnitudes. Combining the predictions from multiple 
networks further improved the scatter. For $i_{\rm AB} = 22.5$ and without 
quality cuts, we found $\sigma_{68}$ to be 50\% lower with \textsc{Deepz}
compared to \textsc{BCNz2}, while the strict outlier fraction
($|z_p - z_s| > 0.02$) reduces from 17 to 10 percent.

This paper tested transfer learning using two different simulations. The
simulation based on the \textsc{FSPS} code performed significantly
better than a template based simulation, indicating the SED modelling 
being important. For both simulations, the photo-$z$ continued improving 
until reaching the maximum number of 
observed redshifts available. This indicated there is further room to improve 
the PAUS photo-$z$ precision. Furthermore, the redshift
performance was shown to depend on the number of
galaxies for different redshifts (Fig. \ref{missing_zbins}).
For high densities, the network is clearly 
superior, but the  template fitting code performs better at redshifts with 
very few spectra. Pretraining with simulations eases the situation, but not 
fully and this is an area of ongoing investigation.

Galaxy surveys typically take multiple exposures in each band, which are
then combined into a single statistically optimal measurement (coadd). 
Since the coadd combines multiple measurements, it can be sensitive to
outliers. We tested methods to include information from individual
flux measurements. Instead of modifying the network architecture, we 
trained the network using coadds generated on the fly from a random
selection of individual exposures. This approach resulted in a 20\% 
reduction in the photo-$z$ scatter (Fig. \ref{alpha}). Combining
multiple networks led to an additional 10\% improvement.

The network architecture also included an autoencoder, which is useful to
extract features and reducing noise. An autoencoder consists
of an encoder network compressing the input to a set of ten features, while
the decoder network attempts to reconstruct the original input. Optimising
the difference between the input and reconstructed values is known to
reduce the noise. We found a 50-70\% reduction in the errors, with the largest
effect for the blue bands. Furthermore, we showed how the autoencoder can
lead to correlated errors between bands. Including features extracted
from the autoencoder leads to a moderate reduction in the photo-$z$ scatter. The
autoencoder is expected to be more important for the wider fields, since this type of network can be trained without spectroscopic redshifts.

The \textsc{Deepz} code estimates redshift probability distributions (PDF), 
which are not provided by many machine learning codes. The probability
distributions were estimated using a mixture density network (MDN). We
validated the PDFs with the probability integral transform (PIT) and
found the $p(z)$ distributions represent the true underlying probability
distributions, with the exception of some outliers. The PDF, when combining 
the networks, performed even better, mostly due to having fewer outliers 
to model. Lastly, we tested quality cuts based on the PDFs and found 
the \textsc{Deepz} photo-$z$ to exceed the PAUS target performance
when selecting the 50\% best galaxies based on a quality cut.

The uniqueness of PAUS is the wide fields, where PAUS 
have observed a total of 47 sq. deg. with 11, 14 and 20
sq.deg. in the 
CFHTls W1, W2 and W3 fields, respectively.
The SNR of PAUS in the wide fields is comparable to the COSMOS fields, although
with fewer exposures
per galaxy and
band ($\sim 5$ exposures in COSMOS and 3 in the wide fields).
The differences between the fields are the broad bands (CFHT Megacam instead of Subaru), the galaxy parameter
(e.g. size, ellipticity) used for the forced photometry from a different parent catalogue and the spectroscopic training set. All fields are calibrated
relative to the SDSS stars. We are currently working
on validating and homogenising the data reduction for the
different fields. Potentially the \textsc{Deepz} network can
be trained on COSMOS and use transfer learning to adapt
to differences between the fields. Also, the extrapolation beyond the spectroscopic subset is
as always uncertain and should be verified using e.g. galaxy cross-correlations \citep{Schneider2006, Newman2008}.
This is work in progress and beyond the current paper.

In this paper we introduced an efficient deep learning technique for high
precision redshift estimation. The network was tested with PAUS, but many 
ideas are not necessarily restricted to narrow-band surveys. Pre-training 
with simulations holds the promise of combining theoretical knowledge 
and empirical data from spectroscopic surveys. Also, the technique of 
randomly constructing coadds should be applicable to large weak lensing 
surveys, including LSST and {\it Euclid}. 

\section*{Acknowledgement}
The authors thank Jacobo Asorey and Malgorzata Siudek for comments.
The PAU Survey is partially supported by MINECO under grants CSD2007-00060, AYA2015-71825, ESP2017-89838, PGC2018-094773, SEV-2016-0588, SEV-2016-0597, and MDM-2015-0509, some of which include ERDF funds from the European Union. IEEC and IFAE are partially funded by the CERCA program of the Generalitat de Catalunya. Funding for PAUS has also been provided by Durham University (via the ERC StG DEGAS-259586), ETH Zurich, Leiden University (via ERC StG ADULT-279396 and Netherlands Organisation for Scientific Research (NWO) Vici grant 639.043.512), University College London and from the European Union's Horizon 2020 research and innovation programme under the grant agreement No 776247 EWC.

The PAU data center is hosted by the Port d'Informaci\'o Cient\'ifica (PIC), maintained through a collaboration of CIEMAT and IFAE, with additional support from Universitat Aut\`onoma de Barcelona and ERDF. We acknowledge the PIC services department team for their support and fruitful discussions.
CosmoHub has been developed by PIC and was partially funded by the "Plan Estatal de Investigaci\'on Cient\'ifica y T\'ecnica y de Innovaci\'on" program of the Spanish government. 

Work at Argonne National Lab is supported by UChicago Argonne LLC,Operator of Argonne National Laboratory (Argonne). Argonne, a U.S. Department of Energy Office of Science Laboratory, is operated under contract no. DE-AC02-06CH11357.
H. Hildebrandt is supported by a Heisenberg grant of the Deutsche Forschungsgemeinschaft (Hi 1495/5-1) as well as an ERC Consolidator Grant (No. 770935).

We gratefully acknowledge the support of NVIDIA Corporation with the donation of the Titan V  GPU used for this research.
Early research for this paper was done at AI Saturdays Barcelona.

\section*{Data availability}
The PAUS observations are currently not publically available, while
the \textsc{deepz} code is available upon request.

\bibliography{ml}{}
\bibliographystyle{mn2e}

\appendix
\section{\textsc{BCNz2} photometric redshift code}
\label{bcnz_description}
In \citet{Eriksen2019} we described the \textsc{BCNz2} photometric 
redshift code. This code was developed to reach good photometric redshift 
precision with PAUS. The code models the galaxy SED as a linear combination 
of templates

\newcommand{\fmodel}{f^{\mathrm{Model}}}

\begin{equation}
\fmodel_{i}[z, {\bm \alpha}] \equiv \sum_{j=1}^n f^j_i(z) {\bm \alpha}_j,
\end{equation}

\noindent
where $f^i_j$ is the model flux for template $j$ in band $i$. The ${\bm
\alpha}$ vector includes the weights of the different SEDs. The estimated redshift 
probability distribution is given by

\begin{align}
p(z) &\propto \exp\left(-0.5 \min_{{\bm \alpha} \geq 0} \chi^2[z, {\bm \alpha}]\right) \\
\end{align}

\noindent
with the $\chi^2$ expression to minimised being defined by

\begin{equation}
\chi^2[z, {\bm \alpha}] \equiv \sum_{i,NB} \left(\frac{\tilde{f_i} - l_i k \fmodel_{i}}{\sigma_i} \right)^2
+ \sum_{i,BB} \left(\frac{\tilde{f_i} - l_i \fmodel_{i}}{\sigma_i} \right)^2.
\label{chi2_eq}
\end{equation}

\noindent
Here the minimisation algorithm \citep{Sha2007} ensures positive amplitudes (${\bm \alpha}$). The
factors $l_i$ are global zero-points per band (i), while $k$ is a free scaling
between narrow and broad bands per galaxy. These factors were introduced to
reduce the sensitivity to calibration problems and issues in the PAUS photometry.

The zero-points $l_i$ were calibrated by comparing the observed flux and
the best fit model when running the photo-$z$ code at the spectroscopic
redshift. This additional zero-point calibration is commonly used and can
account for residuals in the instrumental calibration. However, this method
can effectively adjust the templates, introducing an erroneous zero-point
calibration for a subset of galaxies. We are currently in the
process of building on the work in \citet{Eriksen2019} and have studied
the impact of the additional zero-point calibration (Alarcon in prep.). 
The \textsc{Deepz} code has the advantage of not requiring this calibration
step, 
since it is a machine learning method which directly maps observed
quantities to the redshift.

\section{Effect of photometric outliers}
\label{effect_outliers}
Estimating the photometric redshift with a template fitting code relies on an
analytical likelihood function specifying the data probability given a
model. This is the case for e.g. \textsc{LePhare} \citep{Arnouts2011, 
Ilbert2006} and \textsc{BPZ} \citep{Benitez2000}. In the likelihood and
fitting, the input data are often assumed to have Gaussian and known errors.
Unfortunately, observed data also includes outliers, which are not reflected
in the likelihood. For PAUS, there are problems in the calibration, the
photometry, cross-talk between CCDs and other issues. While
removing outliers is a target of the PAU data reduction, there will always
be some errors remaining.

Ideally the photo-$z$ code should be insensitive to outliers in the input
data. Template fitting codes can in theory be extended to model the outliers
by modelling the flux errors as a linear combination of the standard error and
a wider Gaussian describing the outliers. In practice the idea has multiple
complications. Many photo-$z$ codes rely  on the specific functional form
of the likelihood ($\chi^2$) expression. The \textsc{BCNz2} code use a 
non-negative  minimisation algorithm working with quadratic functions, which 
makes it hard to incorporate many ideas. Furthermore, modelling the outliers would
require setting the outlier rate, which should potentially depend on the 
SNR of the input data. 

\begin{figure}
\xfigure{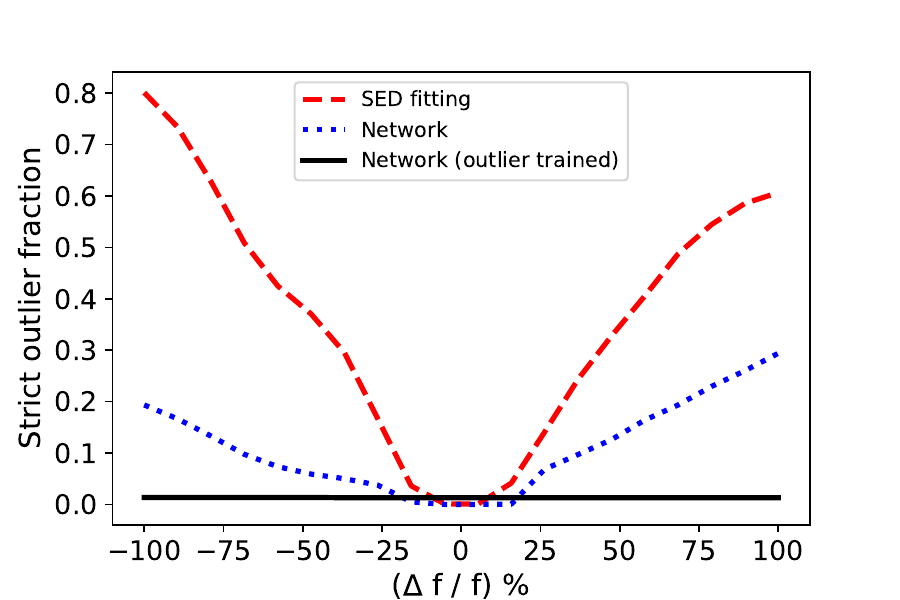}
\caption{The effect of outliers in the photometry on the strict photo-$z$ outlier
fraction. On the x-axis is the percentage $u$-band flux added to all galaxies
in the test set. The result is shown for a SED fitting, a neural network and
a neural net trained with simulations including outliers.}
\label{outlier_frac}
\end{figure}

Machine learning codes are often more robust towards
photometric outliers. To test this idea, we have generated a simple set
of galaxy mocks. In this test, we generate a set of 10000 elliptical 
galaxies. These use 8 elliptical galaxies SEDs without extinction or emission
lines, corresponding to the first template set in \textsc{BCNz2} (run 1). 
We add Gaussian noise with SNR of 10 and 35 in the narrow and broad bands,
respectively. The outliers are generated by adding an
additional flux in the $u$-band to all galaxies in the
test set (see Fig. \ref{outlier_frac}).

Figure \ref{outlier_frac} demonstrates the impact on photometric outliers on
classical template fitting and machine learning approaches. The figure
shows the strict outlier fraction (Eq. \ref{outlier_def}) as a function
of an additional flux fraction applied to the $u$-band for all galaxies
in the test sample. For the "SED fitting" line, we fit the galaxies to
elliptical templates with a minimum $\chi^2$ approach. This is an optimistic
estimate since we perfectly know the SEDs and have not included other
galaxy types. For the network we use the architecture and training procedure 
outlined in \S \ref{data_architecture}.  

The template fitting is strongly sensitive to the outliers, while the neural
network is less sensitive. An approach to further reduce the impact of 
photometry outliers is adding outliers to the training data. In this way
the network learns to not blindly trust features since they could also
be photometric outliers. We have tested adding 10\% outliers uniformly 
distributed over the different bands and with varying amplitudes which is shown in Fig. \ref{outlier_frac}. Note that 
we are not informing the network which galaxies have problematic photometry. 
The network trained with photometry outliers becomes remarkable insensitive
to these, as indicated by the essentially flat solid line.

\section{Deep learning basics}
\label{deep_basics}
Deep learning and artificial intelligence (AI)
has become an important trend over 
the last years. The usage of graphical processor units (GPUs) with massive 
parallelization has enabled training large models with large amounts of data. Furthermore, this 
renewed interest has introduced a new set of different techniques like 
generative adversarial networks (GANs) \citep{Goodfellow2014}, reinforcement
learning \citep{Kaelbling1996}, new network architectures \citep{Kaiming2015}
and the attention mechanism \citep{Vaswani2017}. 
While this paper only uses a small set of techniques, it benefits from
the overall activity in the field. This includes the access to well documented, 
open-source 
libraries for neural networks, like \textsc{PyTorch} 
\citep{Paszke2017} and \textsc{Tensorflow} \citep{tensorflow2015-whitepaper}.

Neural networks are a machine learning technique, which has a long history with 
early implementations in the 1950s \citep{Rosenblatt1958}. The usage of neural
networks was in periods overpromising with successive periods of being out
of fashion. Groundbreaking results on image classifications achieved by 
training a larger neural network with many images led to renewed interest in
the field \citep{Krizhevsky2012}.

Deep learning is effectively a neural network with many \emph{layers}. The network consists of multiple layers or transformations of the data. 
While the performance with a few layers
tends to flatten when increasing the amount of training data, the performance
of deep networks tends to increase with more data. This training is often
computationally expensive but can use graphics processing units (GPUs),
which supports massive multiprocessing. 

There are multiple types of neural networks, including convolutional neural
networks (CNNs) and recursive neural networks. In this paper we use a linear 
neural network and will briefly explain these. The network 
consists of a series of transformations to the data, or \emph{layers}, which 
are sequentially applied to the data. A linear layer is the transformation

\begin{equation}
\rm{linear(}\mathbf{x}\rm{)} \equiv \mathbf{A} \mathbf{x} + \mathbf{b}
\end{equation}

\noindent
where $\mathbf{x}$ is the input data, while the matrix $\mathbf{A}$ and 
vector $\mathbf{b}$ are parameters of the the neural network. These network
parameters will be initiated randomly and trained using the data. 

For the network to learn a non-linear mapping from the input, it 
also need to include a non-linear transformation or \emph{activation function}.
A common choice is the ReLU activation function, which is defined by

\begin{equation}
\rm{ReLU(}\mathbf{x}\rm{)} \equiv \left\{
    \begin{array}{@{}lr@{}}
        0, & \text{for }\mathbf{x} \leq 0 \\
        \mathbf{x}, & \text{for }0\leq \mathbf{x}
        \end{array}\right.
\end{equation}

\noindent
where the operation is performed element-wise. Explained with words, the
ReLU activation sets negative entries to zero. 

In addition, this work uses \emph{Batch Normalisation} \citep{Ioffe2015}.
The batch normalisation is a layer standardising the input to the layer
to have mean zero and unit variance. This transformation is known to 
make neural networks faster to train, more robust and achieve better
performance. When constructing the network, we include a \emph{dropout}
layer, randomly dropping a few percentages of the values. This technique
is often used to hinder \emph{over-fitting}, the 
effect of the network fitting well to the training data, but not generalising
to data for which the network is not trained.

For \emph{supervised training} the network predictions are compared with a
known answer (or \emph{labels}). One then constructs a \emph{loss}, which measures
how wrong the network prediction is. In our case the main contribution to the
loss is given by the negative logarithm of the estimated probability at the
spectroscopic redshifts (Eq. \ref{loss_function}). The training of the 
network is done using \emph{batched}, which is a subset of galaxies jointly
used to estimate the batch loss and update the network. Dividing in
batches is done to train the network faster. When having trained with 
all data once, we say the network has been trained for an \emph{epoch}. 

Updating the network parameters use the \emph{Adam} optimiser. When 
training we include \emph{weight-decay} \citep{Krogh1992}, which is
a technique which adds an additional loss that limits too high parameters.
In practice, this is implemented as a decay term when updating the
weights. Furthermore, how fast the network is updated is controlled by the \emph{learning rate}. A high learning rate reduces the training time, 
but risk the network being stuck in a sub-optimal solution.

\section{Photo-z scatter}
\label{photoz_scatter}
\begin{figure*}
\yfigure{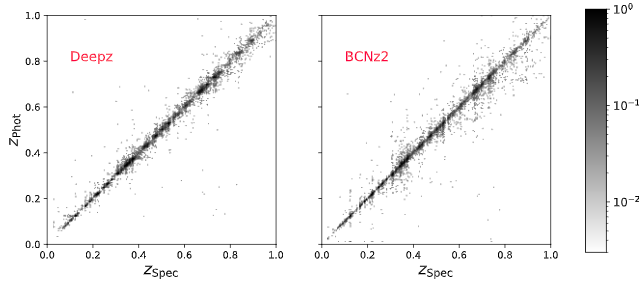}
\caption{Density plot comparing \textsc{Deepz} (left) and \textsc{BCNz2}
(right) redshift predictions to secure zCOSMOS DR3 spectroscopic 
redshifts. The colour scale is logarithmic to view the outliers.}
\label{scatter_plot}
\end{figure*}

Figure \ref{scatter_plot} shows a 2D histogram plot for \textsc{Deepz} and 
\textsc{BCNz2}. The colour scale is logarithmic colour scale to better
visualise the outliers. Otherwise most PAUS galaxies of the galaxies were forming a narrow diagonal line.

\section{Label smoothing}
\label{label_smoothing}
While photo-$z$ estimation is fundamentally a regression problem, it is often
implemented using a classifier (e.g. \citealt{Bonnett2015}), with classes corresponding
to thin redshift bins. For a broad-band survey, one can
typically use bins of $\Delta z = 0.015$. However, for PAUS we need to use
$\Delta z = 0.001$ wide bins to also capture galaxies with excellent 
($\sigma_{68} = 0.001 (1+z))$ redshift precision at the bright end. This would require 15 times more bins. As a result, the size and number of weights in 
the last linear layer, which has a large fraction of the weights, increases
dramatically. The last layer would have approximately 15 times more parameters
for a narrow-band photo-$z$ compared to the broad-band equivalent.

We have tried implementing a photo-$z$ classifier
with different approaches to account for different numbers of galaxies in each class.
In \citet{Buda2017} the authors review the state of
solutions to class imbalance in the literature. They found that oversampling, 
i.e. selecting samples more often from less probable classes, tends to give the best results. 

The classifier approach ignores the information from nearby redshifts. For 
a example, there is no concept of nearby classes when predicting the animal type with a traditional classifier. With PAUS data, the fundamental
limitation is redshift bins without spectroscopic galaxies. When decreasing the bin size, 
there will be bins without galaxies. No matter the weighting scheme, these 
bins will remain empty. Not having a concept of nearby redshift is an 
artefact of framing a regression problem as a 
classification.

One approach to avoid empty redshift bins is label smoothing \citep{Simard2012}. 
Instead of assigning a galaxy to a single redshift bin/class, the redshift is 
randomly scattered to
one of the nearby redshift bins/classes. Applying this technique when 
training significantly reduces the photo-$z$ scatter. Tests with applying
different scatter values, resulted in
a different optimal scatter for bright and faint galaxies, with the 
photo-$z$ scatter reduced most significant for  large redshift scatter.
Instead of using a fixed value or hard-coded relation, we have
developed a method to estimate the required smoothing from the
PDF. In each step, we predict the
$\sigma_{68}$ of the estimated redshift PDF. This can be done fast on the GPU with
a cumulative sum. The redshift scatter is then introduced as a Gaussian smoothing with 15\%
of the $p(z)$ width. While this leads to a significant reduction in the photo-$z$ 
scatter, the resulting photo-$z$ scatter is comparable with an MDN without the
smoothing step. In addition, the PDFs produced by the MDNs were better
giving the true distribution (\S \ref{pz_validate}) and give better
quality cuts. This paper therefore uses the MDN for estimating the
probability distributions.

\section{Network using individual exposures}
\label{all_permutation}
In this paper the default method of including information from
individual exposures is by data augmentations when training the network. This
is done by constructing coadds during the training from randomised subsets
of exposures (see \S \ref{individual_exposures}). Here we briefly report on 
challenges encountered when constructing a network architecture to directly 
estimate the redshift from individual exposures.

A problem for creating a network using individual exposures is the irregularity 
of the data. While galaxy surveys strive to obtain uniform coverage over the 
field, the number of exposures will always depend on the sky-position. The
standard types of neural network work best with regular data, for example
2D images and 1D series of observations. This would require inputting the 
data with all bands having the same number of observations. With observational
data, this would use the lowest number for all galaxies, which would lead
to dropping an unacceptable fraction of observations.

Alternatively, one can construct a larger data structure and set the missing
observations to a special value. Also, the position of the missing values
can be given to the network as an additional mask. These approaches work partly 
for our data, but they do not reach a sufficiently low photo-$z$ scatter. This 
appendix uses a simplified simulation to explain potential pitfalls with 
this approach. For testing the effect of individual exposures, we use the standard network with
one modification. When inputting the individual exposures, the first layer
has the dimension needed for working with a flattened array of the individual
exposures. For simplicity, we did not include the broad-band measurements. 

\begin{figure}
\xfigure{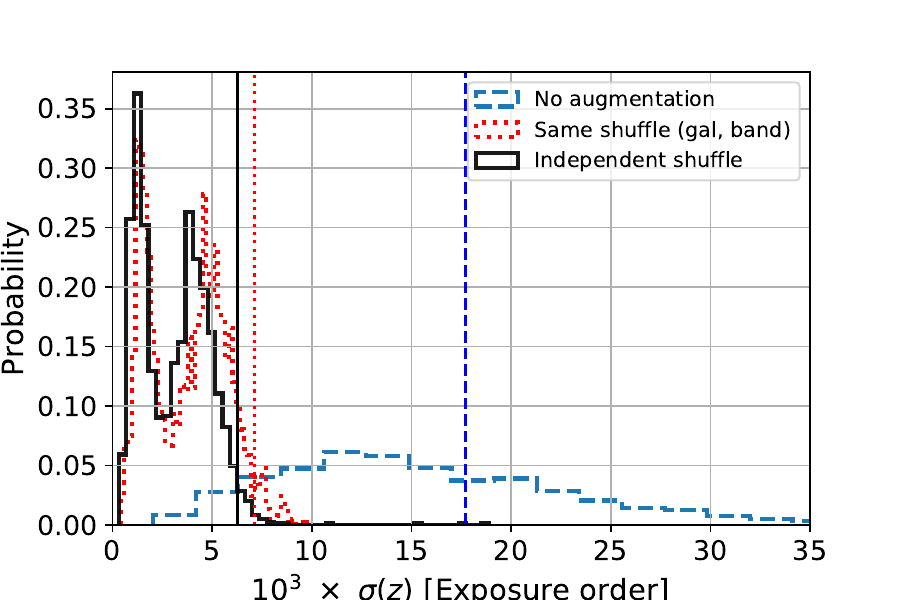}
\caption{The scatter between the photo-$z$ when estimating the redshift with
different exposure orders for different approaches for training
the network (see text). The vertical lines show the $\sigma_{68}$ for a single exposure order.}
\label{ind_fluxes}
\end{figure}

One needs to define an order of the exposures when inputting the individual 
exposures as a matrix (tensor). Given the exposures are a set, 
the network will need to learn
that these measurements have the same meaning. One approach is using
training augmentation, randomly mixing the order of the individual fluxes.
This allows the network to learn more
easily the meaning of the individual
exposures from the limited spectroscopic dataset.

Furthermore, the order also makes a difference when estimating the 
redshifts. Figure \ref{ind_fluxes} shows scatter between photo-$z$ predicted
with different exposures order for different randomisation strategies when 
training the network. Here only 30\% of all individual fluxes are present and the input is given as a dense matrix without
inputting an additional mask. After training the network, we 
test predicting the photo-$z$ for different orders of the exposure. The x-axis 
shows the photo-$z$ scatter between specifying the individual fluxes in a different
order. This is not the normal photo-$z$ error, but an additional error coming
only from the ordering of the individual fluxes.

The "No augmentation" line is trained without any data augmentation, leading to a large 
photo-$z$ scatter. Reordering the measurements during training can be 
computational expensive. One simple approach is to only switch the order of
the input fluxes in each batch. This still produces $5! =120$ orders of the input per band for five exposures. 
As shown in the "Same shuffle" line, this decreases the spread among the predictions. Finally, we have tested fully permuting the different inputs when training. For 
40 bands, this implies ${40}^{120}$ different configurations of which 
we will only sample a small subset. The "Independent shuffle" line shows the 
result when fully randomly selecting the exposure order for each galaxy and band, 
each time the network is trained (epoch). This further improves the scatter, but it's still too large.

\end{document}